# Investigating the use of Software Agents to Reduce The Risk of Undetected Errors in Strategic Spreadsheet Applications

Pat Cleary, Dr David Ball, Mukul Madahar, Simon Thorne,
Christopher Gosling, Karen Fernandez
UWIC, Business School, Colchester Avenue, Cardiff, CF23 7XR, United Kingdom

Pcleary@uwic.ac.uk, Dball@uwic.ac.uk, Mmadahar@uwic.ac.uk, Sthorne@uwic.ac.uk

## 1.0 Introduction

*"To err is human. To really foul things up requires a computer." Anon*

There is an overlooked iceberg of problems in end-user computing. Spreadsheets are developed by people who are very skilled in their main job function, be it finance, procurement, or production planning, but often have had no formal training in spreadsheet use. IT auditors focus on mainstream information systems but regard spreadsheets as user problems, outside their concerns. Internal auditors review processes, but not the tools that support decision-making in these processes.

As highlighted by Systems Modelling Ltd. (2003) the questions that need to be raised are: Are any important decisions made in your company supported by spreadsheets? Have these models been tested or reviewed? Do you have internal standards for spreadsheet development? We all know that people make mistakes. Yet end users and their managers have the confident belief that their work is perfect.

This paper highlights the gaps between risk management and end user awareness in spreadsheet research. In addition the potential benefits of software agent technologies to the management of risk in spreadsheets will be explored.

This paper will also discuss the current research into end user computing and spreadsheet use awareness. The bulk of current research, as below, is taken from two Masters Dissertations completed by Chris Gosling and Karen Fernandez. These two authors completed their research in May 2003 at the university of Wales Institute Cardiff (UWIC). The research focuses on a large NHS trust and Revlon respectively.

## 2.0. The Extent Of Spreadsheet Error

The KPMG survey (Chadwick, 2002) of financial models based on spreadsheets found that 95% of models were found to contain major errors (errors that could affect decisions based on the results of the model), 59% of models were judged to have 'poor' model design, 92% of those that dealt with tax issues had significant tax errors and 75% had significant accounting errors. An article in *New Scientist* (Ward M, 1997) has reported that a study conducted by the British accounting firm Coopers & Lybrand found errors in 90% of the spreadsheets audited. This is an extremely high figure and if the errors went undetected, it could have had a devastating effect on the business.

More broadly, a number of consultants, based on practical experience, have said that 20% to 40% of all spreadsheets contain errors (Panko, 1997). Freeman (1996) cites data from the experience of a consulting firm, Coopers and Lybrand in England, which found that 90% of all spreadsheets with more than 150 rows that it audited contained errors. One Price Waterhouse consultant audited four large spreadsheets and found 128 errors (Ditlea, 1987). According to Franz Hormann (1999) what makes the matters worse is that many spreadsheets are templates, or models, to which users continually add information. If the original contains an error, each new data input amplifies that original error.

Ray Butler of the Computer Audit Unit, Customs and Excise, who, for over ten years, has been investigating errors in spreadsheets used by companies for calculating their VAT payments, says that, "The presence of a spreadsheet application in an accounting system can subvert all the controls in all other parts of that system". On the other hand, David Finch, Head of Internal Audit at Superdrug plc, believes "The use of spreadsheets in business is little like Christmas for children. They are too excited to get on with the game to read or think about the 'rules', which are generally boring and not sexy".

David thinks that we shouldn't be too surprised at the high rate of spreadsheet errors as end user computing is inherently a high risk area. Most aspects, especially those where responsibility clearly lies with the I.T. function, tend to be well controlled as the disciplines surrounding client/user or mainframe environments tend to be transported through to the IT function to provide solutions. However, David believes that control is often much weaker where powerful analysis tools are given to less disciplined users. He is quite frank about what he believes is the cause of the phenomenon "There is often little control over end user developments in spreadsheets with little if any standardisation in development processes by users in different departments, little risk analysis and a general assumption that models, on which important business decisions are made, are accurate. Users who are technically capable of developing applications have not been

## 3.0 Risk Management

As highlighted by Panko (1995), 'an oil and gas company in Dallas lost millions of dollars in an acquisition deal, and several executives were fired. The root of the problem: A spreadsheet model contained an error. The Executives had based their actions on inaccurate spreadsheet data'. The part that is visible is just the tip of the iceberg. The spreadsheets are turning into an integral part of the organisational decision making process. Not only that, they are becoming an integral part of the Information Systems of the organisation, as in many cases data from these spreadsheets is directly fed into the systems and vice-versa. What is needed is a dedicated risk management platform to put a control on the development of such erroneous models. Spreadsheet content in current circumstances is difficult to control and simple errors can lead to costly trading losses. The figure below highlights a basic risk management cycle.

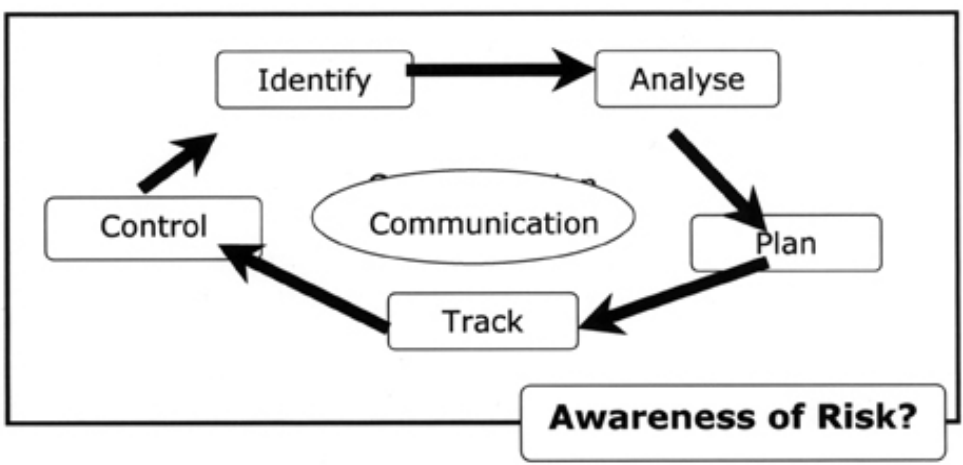

**Figure 1: Risk Management Cycle "The Missing Link"(Madahar, 2003)**

The figure above highlights a simple risk management sequence for any problem. But, the biggest drawback about this chain as far as spreadsheet errors is concerned is "Awareness". It is very hard to formulate a risk management/aversion strategy because one cannot formulate a strategy unless they are aware that spreadsheets are a potential risk to the organisation.

**4.0. Background To Software Agents And Spreadsheets**

There are many proposed solutions to spreadsheet risk management and many different approaches to research on this topic. Software Agent technology has been around since the concept of 'actors' was conceived in the late 1970's by Carl Hewitt's research into Distributed Artificial Intelligence (DAI) (Hewett and Inman, 1991). Actors essentially gave birth to the concept of software agents, having the same principles of manipulating an electronic environment. Hewitt's research was abandoned in the early 80's due to lack of computing power. Due to the recent revolution in computing power, research into software agents has begun a revival of the technologies.

**4.1 Introduction to Agents**

The natural environment of Agent technology is distributed. That is to say Multi Agent Systems (MAS) are in their element in the distributed environment. Agent technologies have had much success in the distributed format on the Internet. Commonly Agent technologies are used for data mining, document retrieval and analysis of information retrieved from the Internet. It is those last two applications that first indicated that agent technologies might be a legitimate solution to risk management in spreadsheets. A MAS could be delivered to company intranet to control and analyse spreadsheet use and content.

**4.2 Agent technologies**

Agent technologies are often described as: software agents, MAS, Agents, and Agent Based Systems. All of these terms essentially describe the same thing. Software Agents that is a MAS, are one or more agents operating in a software environment (see 3.3). Multi agent systems and Agent based systems describe a system, using agents, to solve a problem. Agent technologies are the generic term used to describe any technology that uses agents to function. Croft describes what makes an agent in his paper *'Intelligent software agents - definitions and applications'* (Croft, 1997). Croft identifies the individual characteristics that make up a software agent, as follows:

**Software**

Computer programs developed to accomplish a predetermined task using declarative, procedural or semantic processes.

**Agent**

One who acts on behalf of another, a client, to accomplish a goal. Agents posses the characteristics of *delegacy, competency* and *amenability.*

**Delegacy**

Discretionary authority to autonomously act on behalf of the client. Actions include making decisions, committing resources and performing tasks

**Competency**

The ability to perform given tasks successfully and to manipulate the problem domain environment to accomplish these tasks

**Amenability**

The ability to adapt behaviour in order to optimise the performance in a dynamic environment to accomplish the goals of the client.

### 4.3 Software agents

An Agent that operates in a *software environment.* These include operating systems, computer applications, databases, networks (LAN and WAN) and virtual domains (D.Croft 1997). Agents can also be broken down into types depending on the charteristics that they possess, see Figure 2.

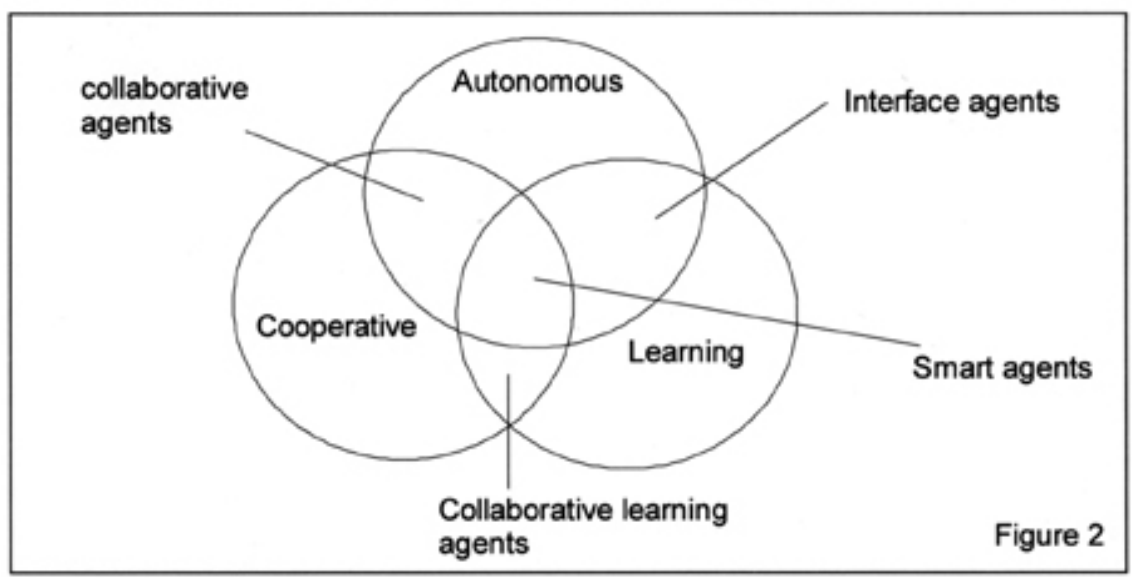

**Figure 2 Agent Typology (Nwana, 1996)**

Nwana's concept of Agent typology differentiates the basic types of agent. Each type is based on the characteristics that it possesses. It is then possible to design a set of Agents to a system's specific needs.

### 4.4 An Introduction to MAS

As previously defined, MAS describes an agent based system using more than one agent. Multi Agent Systems are where Software Agents really come into their element. Agents in a MAS collaborate with each other to achieve the client's goals or the Agent's own goals. A collaborative agent in a MAS communicates with other agents using a complex structured language called Agent Communication Language (ACL). These messages can have a variety of meanings, which range from a request for information from other agents to commanding another agent to undertake an action. Multi agent systems have a specific architecture as defined in figure 3.

This example architecture details MAS as a concept and examines how a MAS can operate. In this example, the client requests information from an Agent (1). The agent will then seek this information, firstly in its 'known' stores and secondly by scanning whatever 'open' information source is available to the agent in its domain. If the agent fails to find the desired information on its own, it will then pass the queries to other agents who will then search for that same information. Assuming that an agent other than the agent with the initial inquiry passes the query, the agent will then return the relevant information to the client and record the 'new' information that was passed to it by the other agent(s). The agent learns new information, so if the same query were to be asked again, the agent could give an immediate response. It is assumed that the agent that initially receives the query is then the co-ordinator for all other agents in each instance.

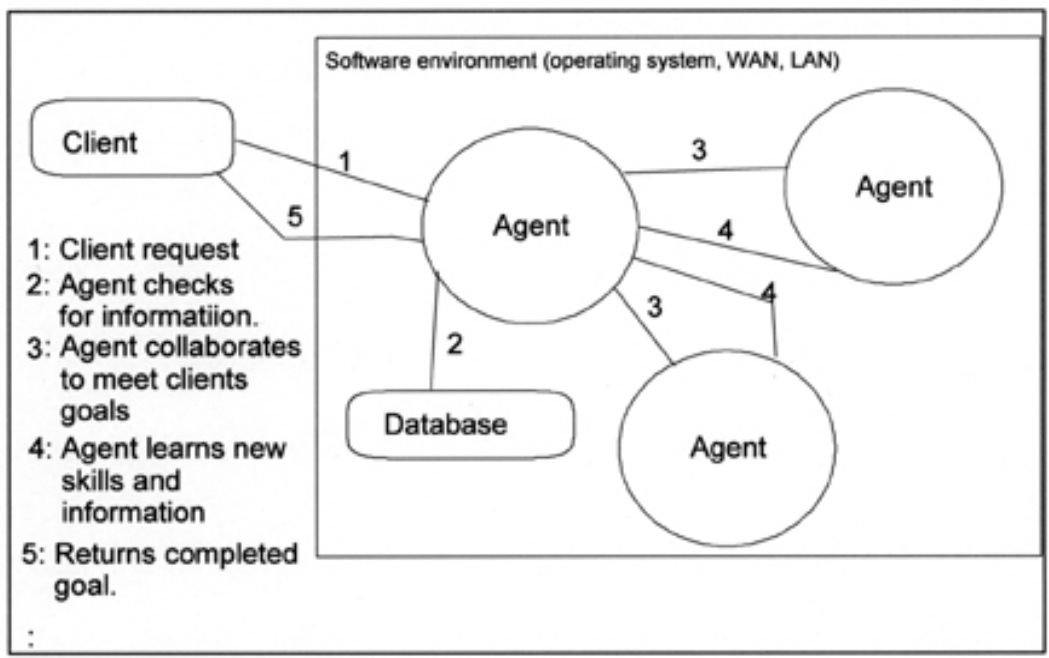

**Figure 3 MAS Relational Model (Thorne, 2002)**

The example in figure 3 is purely a reactive environment, the user invokes the agent, but the principles of the model can be easily applied to a 'proactive' environment. In the context of spreadsheet risk management, the agent would proactively search for spreadsheets and then notify others when it does. If one agent were to find ten spreadsheets in one place the agent would then contact nine other agents that were not already engaged, and then pass the location of the spreadsheet to them.

**5.0. MAS As A Management Solution**

This section explores how a MAS could be delivered to help manage spreadsheet risk. In any organisation the use of spreadsheets as information systems is probably very common. There could be literally hundreds of information systems residing on the organisation's server, networked drives and the hard drives of individual machines. Even if an organisation could realise the risk associated with these 'rogue information systems', auditing and correcting these systems would be a mammoth task costing the organisation time and money. Even when that task is complete the organisation would need to have either some kind of rolling audit system or control system to help manage spreadsheets.

The Multi Agent System that is proposed will research all of these issues including automated correction of mistakes and undesirable behaviour. Intranet Agents have access to every drive, server and networked machine within the organisation. Agents can then identify spreadsheets on the company intranet and classify them, in risk terms, according to criteria derived from research to determine the strategic risks associated with spreadsheet use. The classification process is an inference-based calculation based on the variables extracted from each different spreadsheet, the agent decides how to classify it based on available evidence.

Once classified the agent will analyse the spreadsheet for errors and undesirable or poor practices, such as cells being hidden from the spreadsheet. Once errors are analysed, the agent is able to make appropriate changes or corrections autonomously and a report detailing changes will also be sent to the relevant member of staff to notify what has been changed. As the nature of agents is autonomous, not needing to be summoned like a program or application, they will constantly be searching for spreadsheets on the system twenty four hours a day, thus tackling the 'rolling audit system' needed to ensure that future use of spreadsheets conforms to standards set by controls. The agents could also act as a control for users attempting to develop a spreadsheet, ensuring they classify the spreadsheet correctly and then checking for errors or poor practices.

A great deal of work will be needed in order to construct an effective MAS as a solution to this problem.

**6.0. Current Research**

The Information Integrity Research Centre conducts spreadsheet research at the University of Greenwich as a project. Initial work on spreadsheet errors was first begun in 1996 and finished in June 1997. It involved the researching of spreadsheet creation and use with a view to determining a usable strategy for preventing and detecting spreadsheet errors in business organisations. The team reached the conclusion that what was needed was a strategy involving awareness, prevention and detection that would tend towards zero-tolerance of error-producing situations. The team determined that such a strategy would need to address the following:

- error awareness,
- spreadsheet building methodologies,
- administration/control of spreadsheet developments
- methods of costing errors and audit time.
- teaching and training methods

The following two case studies highlight the issues involved in end user computing and spreadsheet awareness. As mentioned in the introduction, both of these case studies are complied from Masters Dissertations completed in 2003.

**7.0 Case study A**

**7.1 Introduction**

Chris Gosling's paper entitled '*To what extent are systems design and development techniques used in the production of non clinical corporate spreadsheets at a large NHS trust*' examines in detail the use of spreadsheets across a diverse range of departments, from finance to surgery in a large National Health Organisation (NHS) Trust, (Gosling, 2003).

Gosling's aims for the project were:

1. Determine if there is evidence of the application of formal methods in the development of spreadsheets.

2. Determine what relationships there might be between the perceived quality of spreadsheet used and the environment in which they are used.

## 7.2 Methodology

The inspiration for this dissertation was a document produced by the afore mentioned NHS trust entitled 'better information - better health' which outlined the foundations for improved information management and technology at the NHS organisation, an initiative designed to improve IT services within the NHS.

Gosling deployed a Qualitative method of information gathering, distributing questionnaires via electronic mail and hosting the questionnaire on-line to be accessed remotely. The questionnaire was composed of both open and closed questions with a variety of answer types. The questions responses were in several formats including: yes/no; Scaled answers; multiple option answers; free text answers. The questionnaires were distributed amongst all departments via departmental heads of staff.

## 7.3 Findings

The findings of this research were quite profound. In relation to research aim 1, the conclusion was that:

"Gathering 'evidence' of formal methods has proved to be, perhaps the most simple to determine-there appears to be little if not none!" (Gosling, 2003)

Gosling found little to no evidence of formal methodology used to develop spreadsheets and essentially information systems. The general feeling from user groups in relation to aim 2 were that spreadsheets were not an important part of the organisations Information system as illustrated by table 1 below. Table 1 details how subjects responded to the question 'How important are the spreadsheets you develop?' More alarmingly, users were asked about the methodologies deployed to develop spreadsheet systems and the use of macros in those spreadsheets. Tables 2 and 3 detail these results

|  | To You | | To Office | | To Directorate | | To Service Group | | To Trust | |
|---|---|---|---|---|---|---|---|---|---|---|
|  | No. | % | No. | % | No. | % | No. | % | No. | % |
| Not important | 3 | 3% | 1 | 1% | 14 | 14% | 24 | 23% | 29 | 28% |
| Little importance | 1 | 1% | 0 | 0% | 4 | 4% | 5 | 5% | 3 | 3% |
| Some importance | 5 | 5% | 0 | 0% | 4 | 4% | 7 | 7% | 9 | 9% |
| Moderate importance | 17 | 17% | 19 | 18% | 10 | 10% | 6 | 6% | 1 | 1% |
| Important | 49 | 48% | 56 | 54% | 47 | 46% | 14 | 14% | 6 | 6% |
| Very important | 28 | 27% | 27 | 26% | 21 | 20% | 15 | 15% | 10 | 10% |
| Don't Know | 0 | 0% | 0 | 0% | 3 | 3% | 32 | 31% | 45 | 44% |

Table 1 (Gosling, 2003)

**Table 2 Formal methodology applied to spreadsheets? (Gosling, 2003)**

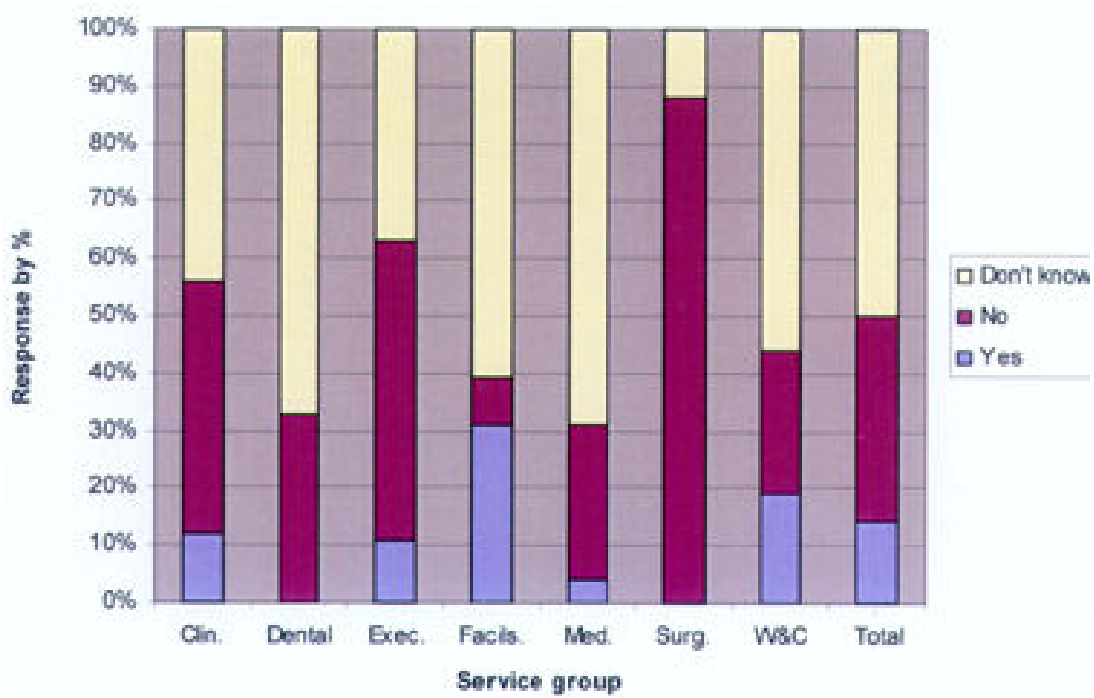

From table 2 the overall conclusion is that 35% did not use a formal methodology to develop spreadsheets and a further 50% being unsure if a formal methodology was used. Only 15% could say that a formal methodology was implemented when developing spreadsheets. This table does not even examine the possibility that the methodologies used were inappropriate, misused or misunderstood. This data helps emphasize the size of the problem faced by some organisations.

**Table 3 Does spreadsheet contain macros?** (Gosling, 2003)

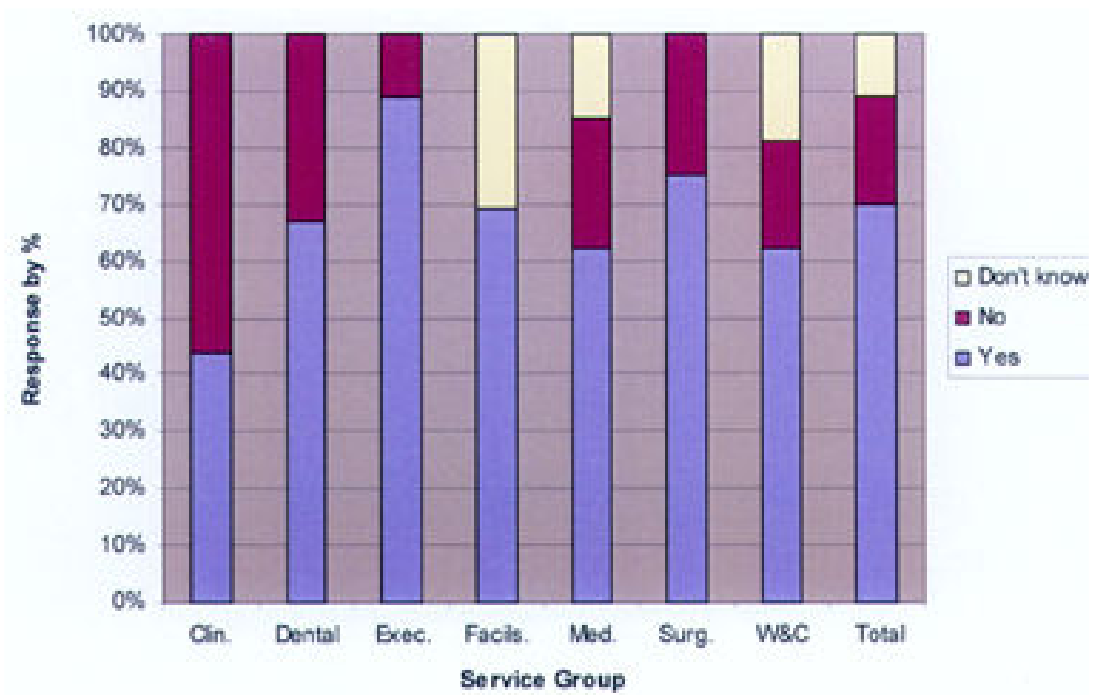

Table 3 details responses concerning the use of macros. The overall results show that 70% of all spreadsheets contained macros. This statistic combined with the lack of a formal methodology to develop the spreadsheet paints an alarming picture, where spreadsheets are developed using varying levels of macro complexity. This indicates the almost complete lack of control over potentially devastatingly inaccurate models.

The over all conclusion that can be drawn from the research is that, like many organisations, the NHS has a major issue with formal methods being deployed for development of spreadsheets and also that users generally have the opinion that spreadsheets they develop are not important to the organisations overall IT operational protocol.

### 8.0 Case Study B

### 8.1 Introduction:

Karen Fernandez's dissertation titled 'Investigation and Management of End User Computing' considers the validity of benefits and disadvantages of EUC (End User Computing) in the Revlon International Corp (UK), the cosmetics company. Karen Fernandez's aim for the study was to compare the findings and conclusions of previous research against the experiences of Revlon's EUC developers. The key objectives of her study were:

- The reasons for growth and advantages EUC can provide
- To investigate the methods used in EUC development, with regard to structured methodology.
- The possible disadvantages and potential 'risks' to the organisation from poor quality EUC applications.
- The training and experience level of the users with regard to EUC tools and IS structured methods in general.
- To assess the perceived need for structured methodology or policy for managing EUC development.

### 8.2 Methodology:

In this study the information gathered related to the views and perceptions of Revlon Staff who are involved in EUC development. Interviews were selected as a method of information gathering. Preliminary interviews with departmental managers were undertaken. Interviews were kept as informal as possible, with reassurances of anonymity and confidentiality given. The discussions were allowed to flow freely in order to gather fuller and wider information.

### 8.3 Findings:

Revlon's perceived usage of EUC was found to be at 51%, albeit this was based on very approximate figures. In general it was found that EUC development is of significant importance to the users, and is used widely throughout the organisation. The departments that deal with strategic/tactical decisions involving 'analysis' tend to be highly dependent on EUC developed applications. All the users consider Excel to be an extremely quick and useful tool, providing them with all the functionality they need. Investigation of the organisation revealed that:

- Feasibility analysis is rarely carried out.
- There is no change control on the EUC applications, except periodic archiving, and is considered sufficient.
- No detailed documentation. For e.g. 'The Finance Department experiences this disadvantage as they have undocumented complex functions'.

- Although their tool skill and knowledge is very high, their system development methods knowledge is very low, and although they believe their applications to be of good quality, the existence of examples of potential risk incidents puts this good quality into question.

The results from this study display that EUC development in Revlon despite the users' perceptions that Revlon is not at risk, there is evidence to the contrary with incidents of errors being found at senior management. According to Fernandez the EUC can be of strategic importance, defining the policy to manage EUC development must be driven by senior management rather than by the users themselves. However, she mentions that for a policy to be most effective it must be supported by the users, who should be convinced of the need for adherence to the policy and believe in the benefits of this would bring.

### 9.0 How Fernandez's and Gosling's research validate that proposed at UWIC

End User Computing is increasingly challenging the dominance of Corporate Computing within organisations. Previous research in the field of EUC has predicted that poor quality and control within EUC applications is putting organisations at great risk. Fernandez's study has considered the validity of these previous findings in Revlon International Corp (UK). By analysing the qualitative information gathered, she stresses that there is the need for a policy or structured methodology for management of EUC in Revlon.

Gosling has gone a step further and focused on Spreadsheets in particular. He has identified the 'large' NHS trust as having a 'trust' wide problem. Although Gosling does not identify the exact or approximate number of spreadsheets in question, his sample was representative of the organisations population. It could also be safely assumed that there are significant numbers to become a serious threat to the organisations functional operations. This is backed up by tables 2 and 3 which detail the lack of formal methodologies deployed.

It is these kind of organisations that UWIC proposes has a strategy for spreadsheet use and implements that strategy due to the extent of risk in the organisation. Furthermore the users indicated in their responses that they might not be aware of the risks associated with spreadsheet use as defined in table 1 of this document.

This research is current and extensive. It clearly indicates that an organisation such as the NHS and Revlon would benefit from strategies and systems as proposed by UWIC.

### 10.0 Aims and Objectives of UWIC Proposed Research:

All in all, the research done to date in spreadsheet development presents a very disturbing picture. Every study that has attempted to measure errors, without exception, has found them at rates that would be unacceptable in any organisation. These error rates, furthermore, are completely consistent with error rates found in other human activities.

We at UWIC as a research group aim to:

"Investigate the means to raise awareness amongst organisations about the importance and the impact of errors in spreadsheets and investigate how Agent Technology can be used to monitor/rectify the occurrence of such errors."

This aim shall be accomplished with the help of following objectives:

1. To identify the possible risks involved in spreadsheets.

2. To classify/categorise these errors in spreadsheets.

3. To formulate a risk management strategy for each category of risks.

4. To implement the formulated strategy

5. To monitor the effects of the strategy

6. Create a specification(s) for agent design based on the strategy (ies) formulated for the reduction of risk in categories of spreadsheet use.

7. Amend that specification in line with current technological capabilities

8. Develop agents using existing technology to meet the above specification(s)

9. Deploy the agents in a test environment

10. Develop means of monitoring the effectiveness of the agents

11. Modify agent design in the light of experience

This research shall proceed with a quasi-experimental approach as the environment of analysis is a natural environment and has no direct intervention from the investigator in the natural occurrences. The author shall investigate the proposed plan and the most appropriate methodology during the course of research and subsequently highlight the limitations of the methodologies used. The ecological validity shall be prime concern of the researcher. The researcher shall concentrate on the external as well as internal validity. The reason being that the author intends to find a solution, which, hopefully, should be acceptable and implementable in different organisations.